\documentclass[a4,twocolumn]{article}
\usepackage{graphicx}
\usepackage{graphicx}
\usepackage{amsmath}
\usepackage{amssymb}
\usepackage{booktabs}
\usepackage{cite}
\usepackage{multirow}
\usepackage[strings]{underscore}
\usepackage[pagebackref=true,breaklinks=true,letterpaper=false,colorlinks,bookmarks=false]{hyperref}
\begin{document}
\title{The Unreasonable Effectiveness of Encoder-Decoder Networks for Retinal Vessel Segmentation}

\author{Bj{\"o}rn Browatzki$^{1,2,3}$ \and J{\"o}rn-Philipp Lies$^{1,2}$ \and Christian Wallraven*$^{1,3}$}
\date{%
	$^1$Eye2you GmbH, T{\"u}bingen, Germany\\%
	$^2$University Eye Hospital T{\"u}bingen, Germany\\
	$^3$Dept. of Artificial Intelligence, Korea University, Seoul, South Korea
}

\maketitle              
\begin{abstract}
We propose an encoder-decoder framework for the segmentation of blood vessels in retinal images that relies on the extraction of large-scale patches at multiple image-scales during training. Experiments on three fundus image datasets demonstrate that this approach achieves state-of-the-art results and can be implemented using a simple and efficient fully-convolutional network with a parameter count of less than 0.8M. Furthermore, we show that this framework - called VLight - avoids overfitting to specific training images and generalizes well across different datasets, which makes it highly suitable for real-world applications where robustness, accuracy as well as low inference time on high-resolution fundus images is required.

\end{abstract}
%
%
%
\section{Introduction}
The analysis of retinal blood vessel structure from fundus photographs can help to identify eye diseases such as diabetic retinopathy, macular degeneration, or hypertension at early stages \cite{Abramoff2010}. Often the growth of new vessels, increased tortuosity, and other morphological changes can be observed in patients \cite{Hubbard1999, Cheung2011}. 
A second application domain is the use of the unique vascular structure as a landmark feature for tasks such as biometric authentication \cite{Marino2006} or registration of multiple images of the same eye across visits to track retinal changes over time or align multiple views to obtain a larger field-of-view \cite{Tian2010, Gharabaghi2013, Hervella2018}.

However, accurate tracing of the fine vascular structure is a difficult, labor-intensive and time consuming task. Numerous methods for automatic vessel segmentation have been proposed in the past. These range from early systems relying on low-level features analyzing local image statistics such as Gabor filters \cite{Soares2006} or ridge based features \cite{Staal2004} in combination with various types of classifiers such as support vector machines \cite{Ricci2007, You2011}, k-NN \cite{Staal2004}, boosting \cite{Fraz2012}, or neural networks \cite{Marin2011}.  
In recent years, approaches relying on powerful convolutional neural networks have gained popularity and have led to significant increases in segmentation performance \cite{Fu2016, Liskowski2016}.  

Many of these deep learning approaches follow a patch-based approach in which the fundus image is split into a set of small patches and each patch is processed individually \cite{Liskowski2016, Li2016, Oliveira2018, Jin2019}. These methods offer high performance in terms of pixelwise accuracy but are unable to capture semantic and contextual information as only a small image neighborhood is observed. In addition, these approaches are computationally expensive even for low-resolution images since a high number of patches has to be processed.

In contrast, several methods have been proposed that take the entire fundus image as input and produce the segmentation result in one pass \cite{Fu2016, Son2017, Maninis2016}. These methods typically operate much faster and are also able to produce high quality segmentation masks. Another advantage of this approach is that image-wide knowledge about the shape of the vessel structure across the retina becomes available and can be even incorporated explicitly in the model as, for example, in \cite{Son2017}. A core limitation of this approach, however, is that they almost always also learn the specific "appearance" of a particular training set given that public datasets typically only contain a low number of fundus images. Therefore these methods struggle to generalize to new images from different sources, as retinal images may vary drastically in terms of image resolution, field-of-view, illumination, sharpness, and recording artefacts.

In the present paper, we propose a hybrid approach in order to bridge the gap between local patch-based and global image-based methods. Specifically, we make use of large patches that are extracted at multiple image scales. Our experiments show that this combination is able to  achieve state-of-the-art segmentation performance combined with excellent generalizability across different types of fundus images. 

Importantly, we also show that accurate vessel segmentation does not require large and computationally expensive models. By employing a ResNet-based \cite{He2015} encoder-decoder architecture and incorporating building blocks of modern efficient networks \cite{Howard2017, Sandler2018, Shi2016}, we show that it is possible to downscale model complexity and processing time by an order of magnitude without a loss in performance, culminating in our VLight architecture.

Source code and pre-trained models are available at \url{https://github.com/browatbn2/VLight}.

\section{Studying patch size and model architecture}

In the following section we want to first shed light on how the choice of patch size in training and test affects segmentation performance. Next, we will study how much computational complexity is required to achieve high quality results.
\subsection{Training setup}
\subsubsection{Datasets}
As test bed for our evaluations serve three common fundus image datasets.

The \emph{DRIVE} dataset \cite{Staal2004} is currently the most widely used benchmark in this field. It contains 40 RGB fundus images of size $565 \times 584$ pixels, split into 20 images for training and 20 images for testing. Annotations of the first human observer serve as ground truth. 

The \emph{CHASE\_DB1} dataset contains only 28 images but of higher resolution ($999 \times 964$px) than DRIVE. Unfortunately, there is no fixed training/test split and choices are inconsistent throughout the literature. We follow the majority of the recent methods we compare against \cite{Yan2018, Li2016, Zhuang2018, Alom2019} by selecting the first 20 images as training images and use the remaining 8 images for testing.

The \emph{HRF} dataset contains 45 high-resolution ($3504\times2336$px) fundus images split into 3 sets of 15 images from healthy patients, patients with diabetic retinopathy, and patients with glaucoma \cite{Kohler2013}. We follow Orlando et al. \cite{Orlando2017} using the first five images in each set (healthy, glaucoma, diabetic retinopathy) for training and the remaining 30 images for testing.

\subsubsection{Training procedure}
We do not employ any preprocessing or post-processing steps. We apply common data augmentations of rotation  ($\pm 60^\circ$), horiz./vert. flipping as well as RGB color ($\pm20$), brightness/contrast ($\pm50\%$), and gamma ($\pm20\%$) shifts. All models are trained for 100k samples where one sample constitutes a dynamically selected random crop. 

We use the Adam optimizer \cite{Kingma2014} ($\beta_1=0.9, \beta_2=0.999$) with a learning rate of 0.001 (reduced to 0.0002 after 80k samples) and a batch size of 10. We minimize the binary cross entropy loss between sample ground truth segmentations and predicted vessel probabilities. 

\subsubsection{Evaluation Metrics}

We evaluate the quality of vessel predictions by reporting Area Under Curve for Receiver Operating Characteristic (ROC), Area Under Curve for the Precision-Recall Curve (PR), and F1 measure (or Dice coefficient). For binarized results we also report accuracy (ACC).  Vessel probability maps are converted to binary masks by applying a fixed threshold of 0.5. 
Please note that segmentations contain many more non-vessel pixels than vessel pixels. Accuracy is therefore a less suitable measure for performance in comparison to threshold-free metrics like Precision-Recall or ROC.

As is common practice, we only consider pixels inside the field-of-view for calculating the above metrics and discard black borders surrounding the visible part of the retina. Field-of-view masks are supplied by DRIVE and were determined for CHASE\_DB1 by thresholding median-filtered grayscale images.

\subsection{Effective patch sizes}
Methods that operate on image patches for retinal vessel segmentation use small patches of, for example, $48\times48$ pixels \cite{Jin2019}. These patches only cover a local region of the typically very large fundus images lacking larger-scale contextual information.   
To study the influence of the selected patches on segmentation performance we train a baseline U-Net model \cite{Ronneberger2015} with different patch selection strategies for training and test. We opt for the U-Net model since it has become immensely popular in recent years for medical image segmentation. Numerous methods have been propose for vessel segmentation that build upon the basic U-Net architecture by adding, for example, residual and recurrent blocks \cite{Alom2019}, deformable convolutions \cite{Jin2019}, or efficient bottleneck blocks \cite{Laibacher2018}.
We compare against these methods explicitly in Sec.~\ref{sec:sota}.

\begin{table*}
	\begin{center}
		\vspace{-0.3cm}
        \begin{tabular}{l c c c c c | l l}
			\toprule
            \bf Method            & \bf Patch-size  & \bf Scale    & \bf Patches & \bf Time & \bf F1    & \bf F1          & \bf F1\\
                                  &                 &\bf train/test& \bf (test)  & \bf GPU  & \bf DRIVE & \bf CHASE & \bf HRF \\
			\toprule
			\midrule
            U-Net                        & 544x544 & 1/1 & 1 & \bf 0.04s & 0.8305 & 0.7724 & 0.7478\\
			\midrule
			\midrule
            U-Net \cite{Jin2019}         & 48x48  & 1/1 & 14916  & 3.1s  & 0.8175 & - & - \\
            U-Net                        & 48x48  & 1/1 & 342    & 1.65s & 0.8172 & 0.5290 & 0.4809\\
			\midrule
            U-Net                        & 96x96  & 1/1 & 64     & 0.32s  & 0.8281 & 0.5355 & 0.3935 \\
			\midrule
            U-Net                           & 128x128 & 1/1 & 36  & 0.25s & 0.8304 & 0.5192 & 0.3831\\
			\midrule
            U-Net                           & 256x256 & 1/1 & 9   & 0.08s & \bf 0.8316 & \multirow{2}{0.4cm}{0.4824} & \multirow{2}{0.4cm}{0.2607}\\
            U-Net                           & 256x256 & 1/2 & 25  & 0.25s & 0.7253 & \\
			\midrule

            U-Net                           & 512x512 & 2/2 & 9  & 0.295s & 0.8302 & 0.7722 & 0.7704\\
            \midrule
            \midrule
            U-Net                      & 256x256  & [1-2] / 1 &      9 &   0.08s &  0.8173  & \multirow{3}{0.4cm}{0.7125} & \multirow{3}{0.4cm}{0.7561} \\
            U-Net                      & 256x256  & [1-2] / 2 &     36 &   0.28s &  0.8248 \\
            U-Net                      & 256x256  & [1-2] / 1+2 &   45 &   1.96s &  0.8261\\
			\midrule
            U-Net                      & 512x512  & [2-4] / 2 &      9 &   0.29s &  0.8283 &\multirow{4}{0.2cm}{\bf0.7745} & \multirow{4}{0.2cm}{\bf 0.7954} \\
            U-Net                      & 512x512  & [2-4] / 3 &     16 &   0.51s &  0.8190 &\\
            U-Net                      & 512x512  & [2-4] / 4 &     36 &   1.13  &  0.8273 &\\
            U-Net                      & 512x512  & [2-4] / 2+3+4 & 61 &   1.96s &  0.8299 &\\

			\bottomrule
		\end{tabular}	
	\end{center}
	\vspace{-0.2cm}
	\caption{\small Evaluating the effect of different patch sizes within and across datasets.}	
	\label{tab:patchsize}
	\vspace{-0.6cm}
\end{table*}

Results are listed in Tab.~\ref{tab:patchsize}. We distinguish between three types of approaches here: Full images as inputs (equivalent to a single patch of size $544\times544$px), fixed size patches, and varying patch sizes. We make three important observations here: First, we see that models trained with large fixed size patches produce excellent results when tested on the same dataset (DRIVE) but do not generalize well to images from other datasets (CHASE\_DB1 and HRF). Second, multi-scale testing leads to a significant increase in accuracy if trained on varying resolutions. Lastly, processing time decreases considerably as patch sizes are increased. 
 
\subsection{Efficient architecture}

As studied in the previous section, we can obtain a highly accurate segmentation results by employing a vanilla U-Net architecture. However, the model size and computational costs make it unsuitable for mobile or embedded devices. Our goal in this section is to show that accuracy can be maintained (or even increased) while at the same time reducing processing time by an order of magnitude.

\begin{table*}
	\begin{center}
		\begin{tabular}{l  c  c c c  c }
			\toprule
            \bf Method            & \bf Patch-size  & \bf Scale & \bf Params &  \bf Time  & \bf F1 \\
			\toprule
            U-Net                     & 512x512  & [2-4] / 2+3+4 & 31.03M  & 1.96s &  0.8299 \\
            \midrule
            Simple Baseline (SB) \cite{Xiao2018} & 512x512  & [2-4] / 2+3+4 & 18.96M  & 1.03s &  \bf 0.8312 \\
       SB + Bilinear & 512x512  & [2-4] / 2+3+4 & 16.39M  &  1.21s & 0.8311 \\
       SB + Pixel Shuffle \cite{Shi2016} & 512x512  & [2-4] / 2+3+4 & 13.45M  & 0.83s &  0.8309 \\
       SB + Pixel Shuffle + DWC \cite{Chollet2017} & 512x512  & [2-4] / 2+3+4 & 1.83M  &  0.72s &  0.8307 \\
            \midrule
      VLight        & 512x512  & [2-4] / 2 & 0.74M  & \bf 0.08s &   0.8184 \\
      VLight        & 512x512  & [2-4] / 3 & 0.74M  & 0.15s &   0.8276 \\
      VLight        & 512x512  & [2-4] / 4 & 0.74M  & 0.34s &   0.8261 \\
      VLight        & 512x512  & [2-4] / 2+3+4 & 0.74M  & 0.56s &  0.8299 \\
			\bottomrule
		\end{tabular}	
	\end{center}
	\vspace{-0.2cm}
	\caption{\small Comparison between different model architectures on DRIVE.}	
	\label{tab:arch}
	\vspace{-0.7cm}
\end{table*}
While leaving training and testing procedures untouched, we replace the U-Net with a ResNet-based encoder-decoder model similar to the Simple Baseline architecture of \cite{Xiao2018}. This model consists of the first 3 ResNet layers of a standard ResNet-18 architecture and an inverted counterpart in which 2-strided convolutions are replaced by $4 \times 4$ deconvolutions.

\begin{figure*}
    \begin{center}
        \includegraphics[width=0.9\linewidth]{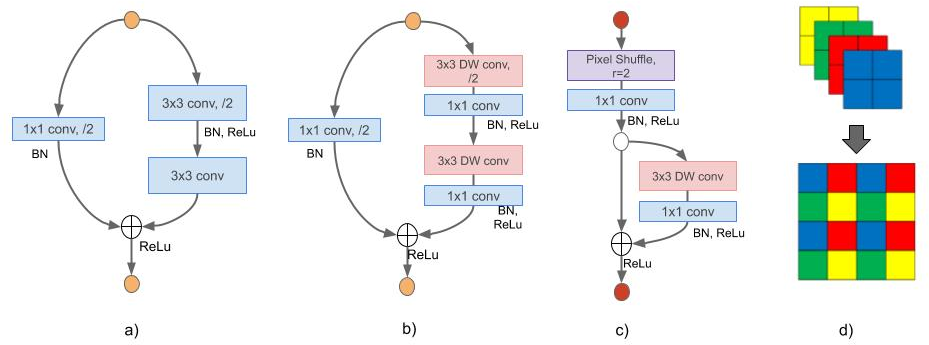}
    \end{center}
    \vspace{-0.8cm}
    \caption{\small VLight network blocks. a) Original ResNet downsampling block containing two convolutional layers; b) Residual downsampling block with depthwise convolutional layers; c) Residual upsampling block with Pixel Shuffle; d) Upscaling using Pixel Shuffle operation}
    \label{fig:blocks}
\vspace{-0.5cm}
\end{figure*}

A common approach for activation upsizing in a decoder pipeline consists in simple bilinear upscaling. In our experiments this, however, led to decreased accuracy and no increase in runtime performance. Shi et al. \cite{Shi2016} proposed an alterative upscaling method in the context of image super-resolution which has also been used successfully employed for semantic segmentation tasks \cite{Cai2019}. Their so-called Pixel Shuffle operation transforms a $H\times W \times C r^2$ tensor into a $rH\times rW \times C$ tensor by simply rearranging element positions as shown in Fig~\ref{fig:blocks} d). This efficient operation leads a reduction in decoder parameters and a processing speed-up without suffering a significant loss in performance. 

\begin{figure*}
    \begin{center}
        \includegraphics[width=0.9\linewidth]{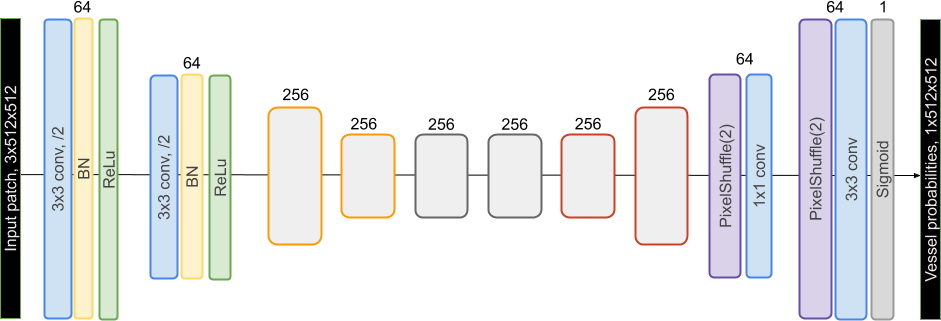}
    \end{center}
    \vspace{-0.6cm}
    \caption{\small VLight network overview. Numbers on top of layers and blocks denote number of output channels.}
    \label{fig:arch}
\vspace{-0.5cm}
\end{figure*}

Next, we make use depthwise separable convolutions (DWCs), introduced in Xception \cite{Chollet2017}, and replace ResNet blocks with blocks akin to MobileNet \cite{Howard2017, Sandler2018}. See Fig~\ref{fig:arch} and Fig~\ref{fig:blocks} for more details. A similar approach is also used by M2U-Net \cite{Laibacher2018} to create an extremely efficient segmentation model for embedded devices. We observe a drastic reduction in parameter count while only noticing a small decrease in accuracy. 

For our final model, VLight, we replace the costly first $7 \times 7$ convolution followed by maxpooling with two $3 \times 3$ convolutions and trade model depth for model width. This tradeoff has been motivated by recent studies, for example in \cite{Wu2019}. We half the number of residual blocks and the set all channel sizes to 256. 

\section{Comparison with state-of-the-art}	\label{sec:sota} 
We first compare our approach with other methods on two low-resolution datasets that are commonly used for the evaluation of vessel segmentation algorithms. In addition to the already used DRIVE dataset, we also test on the CHASE\_DB1 dataset. 

\subsection{Results}

Table~\ref{tab:sota} shows the performance of our approach in comparison to recent state-of-the-art methods for retinal vessel segmentation on the DRIVE and CHASE\_DB1 datasets. 
\begin{table*}
    \footnotesize
	\begin{center}
		\begin{tabular}{l | c |  c c c c | c c c c}
			\toprule
                         \multicolumn{2}{c}{}      &  \multicolumn{4}{c}{\bfseries DRIVE} &   \multicolumn{4}{c}{\bfseries CHASE\_DB1}\\
            \bf Method                           & Year & F1    & ACC    & ROC    & PR      & F1     & ACC    & ROC    & PR \\
			\toprule
            \midrule
            Li \cite{Li2016}                      & 2015 &   - & 0.9527 & 0.9738 & -       &  -     & 0.9581 & 0.9716 & -  \\
            DRIU     \cite{Maninis2016}          & 2016 & 0.8210 & 0.9541 & 0.9793 & 0.9064 & -      & -      & -      & -      \\
            V-GAN     \cite{Son2017}             & 2017 & 0.8277 & 0.9560 & 0.9803 & 0.9142 & -      & -      & -      & -      \\  
            DUNet     \cite{Jin2019}             & 2018 & 0.8237 & 0.9566 & 0.9802 & -      & -      & 0.9610 & 0.9804 & -      \\
            M2U-Net   \cite{Laibacher2018}       & 2018 & 0.8091 &\bf 0.9630 & 0.9714 & -      & 0.8006 & \bf 0.9703 & 0.9666 & -      \\
            R2U-Net \cite{Alom2019}              & 2018 & 0.8171 & 0.9556 & 0.9784 & -      & 0.7928 & 0.9634 &  0.9815 & -  \\
            Yan \cite{Yan2018}                   & 2018 & -      & 0.9542 & 0.9752 & -      & -       & 0.9610 & 0.9781 & - \\
            LadderNet \cite{Zhuang2018}          & 2019 & 0.8202 & 0.9561 & 0.9703 & -      & 0.8031 & 0.9656 & 0.9839 & - \\
			\midrule
            VLight                                & 2020 & \bf 0.8299 & 0.9565 & \bf 0.9820 & \bf 0.9168  &  \bf 0.8194  & 0.9671  & \bf 0.9873 & \bf 0.9057  \\
			\bottomrule
		\end{tabular}	
	\end{center}

	\caption{\small Comparison against other recent approaches on DRIVE and CHASE\_DB1.}	
	\label{tab:sota}
\end{table*}
VLight achieves best non-binarized results on both datasets, outperforming much more complex and slower methods. Only M2U-Net \cite{Laibacher2018} reports higher accuracy - however at considerably lower F1/ROC/PR rates. It is unclear, whether this stems from differences in training loss (M2U-Net is trained on a combination of binary cross-entropy loss and additional Jaccard index loss) or a different thresholding method. Our fixed threshold of 0.5 does not give the highest possible accuracy. However, we refrain from optimizing this threshold since it easily results in overfitting the (small) test set and may lead to less robust results in practice.

\subsection{High-resolution fundus images}
Table~\ref{tab:HRF} summarises segmentation results for our and other methods on the HRF dataset that contains high-resolution fundus images of $3504\times2336$ pixel. On HRF we test only using the original image scale and use Otsu thresholding \cite{Otsu1979} to obtain binary masks. Our efficient model outperforms the previous top-performing method (DUNet) while processing the high-resolution fundus images an order of magnitude faster. M2U-Net offers faster runtimes at, however, significantly reduced performance.

\begin{table*}
	\begin{center}
		\begin{tabular}{l | c | c |  r  r r  r  r r}
			\toprule
            \bf Method                      & Year & Time   & F1      & SE     & SP     & ACC    & ROC & PR  \\
			\toprule
			\midrule
            Orlando \cite{Orlando2017}      & 2017 & 5.8s$^*$   & 0.7158  & 0.7874 & 0.9584 & -      & - & - \\
            Yan et al. \cite{Yan2018}       & 2018 & -      & 0.7212  & \bf 0.7881 & 0.9592 & 0.9437 & - & - \\
            M2U-Net \cite{Laibacher2018}    & 2018 &\bf 19.9ms$^\dag$& 0.7814& -      & -      & 0.9635 & - & - \\
            DUNet \cite{Laibacher2018}      & 2018 & - [$>$1min]$^\ddag$& -   & 0.7464 & \bf 0.9874 & 0.9651 & 0.9831 & - \\
			\midrule
            VLight                            & 2020 & 460ms$^\P$  &\bf 0.8145 & 0.7537 & 0.9873 & \bf 0.9656 &\bf 0.9850 & \bf 0.8977 \\
			\bottomrule
		\end{tabular}	
	\end{center}
    \caption{\small Evaluation of segmentation results and runtime performance on the high-res. ($3504\times2336$px) HRF dataset. 
    $^*$Intel Xeon E5-2690, $^\dag$$^\ddag$GTX 1080Ti, $^\P$TITAN X.
    }	
	\label{tab:HRF}
\end{table*}

\subsection{Cross-dataset evaluation}
Due to the high human effort involved in creating ground truth annotations for retinal vessel segmentations, available datasets are very small. DRIVE, CHASE\_DB1, and HRF only contain 40, 28, and 45 images respectively. Modern deep networks are easily able to fit closely to the data distribution of the training set which poses a high risk of overfitting. Furthermore, there exists a clear danger of test set overfitting through repeated hyper-parameter tuning.

\begin{table*}
    \footnotesize
    \begin{center}
        \begin{tabular}{l  c | c c c | c c c | c c c
            }
            \toprule
                                    &        &  \multicolumn{3}{c}{\bfseries DRIVE}  &  \multicolumn{3}{c}{\bfseries CHASE}  & \multicolumn{3}{c}{\bfseries HRF} \\
            \bf Metric              & \bf Method                    &  D & C & H     &   D & C & H &    D & C & H  \\
            \toprule
            \toprule
            \multirow{4}{0.5cm}{F1} & ErrorNet \cite{Tajbakhsh2020} & -      &  -    & -  &   73.20 & 81.5  & 68.6    & - & - & - \\ 
                                    & V-GAN \cite{Son2017}          & -      &  -    & -  &   69.30 & 79.7  & 66.4    & - & - & - \\
                                    & U-Net \cite{Zhuang2019}       & -      & 50.69 & -  &   65.05 & -     & -       & - & - & - \\   
                                    & AMCD \cite{Zhuang2019}        &  82.33 & 73.95 & -  &   78.60 & 80.15 & - \\   
             \cmidrule{2-11}
                                    & VLight                          &  \bf 82.84 & \bf 76.00 & \bf 79.21 &    \bf 81.65 & \bf 81.94 & \bf 81.31  & \bf 80.31 & \bf 75.37 & \bf 81.45\\  
            \toprule
            \toprule
            \multirow{5}{0.5cm}{ROC} & Li \cite{Li2016}             & 97.38 & 96.28  & -  &    96.05 & 97.16 & -      & - & - & - \\ 
                                     & U-Net \cite{Zhuang2019}      &  -    & 68.89  & -  &    92.18 & -     & -      & - & - & - \\   
                                     & AMCD \cite{Zhuang2019}       &  97.77 & 96.43 & -  &    96.91 & 97.85 & -      & - & - & - \\   
             \cmidrule{2-11}
                                    & VLight                          &  \bf 98.12 & \bf 97.70 & \bf 98.10   & \bf 97.80 & \bf 98.73 & \bf 98.44 &      \bf 97.16 &\bf 96.89 &\bf 98.50 \\  
            \toprule
            \bottomrule
        \end{tabular}	
    \end{center}
    \caption{\small Evaluation of cross-dataset segmentation performance to study generalizability of our and other methods. D=DRIVE, C=CHASE\_DB1, H=HRF.}
    \label{tab:crossdb}
\end{table*}

To study the robustness and generalizability of our method, we evaluate segmentation performance across datasets. We use our previously trained models and predict vessels on another unseen dataset without any adaptation to the new dataset. Tab.~\ref{tab:crossdb} shows that our multi-scale training scheme leads to a high degree of adaptability to the different images sizes and appearance across the three tested datasets. Our simple and efficient approach even outperforms ADCM, a method explicitly dedicated to domain adaptation. 

\begin{figure*}
    \begin{center}
        \includegraphics[width=\linewidth]{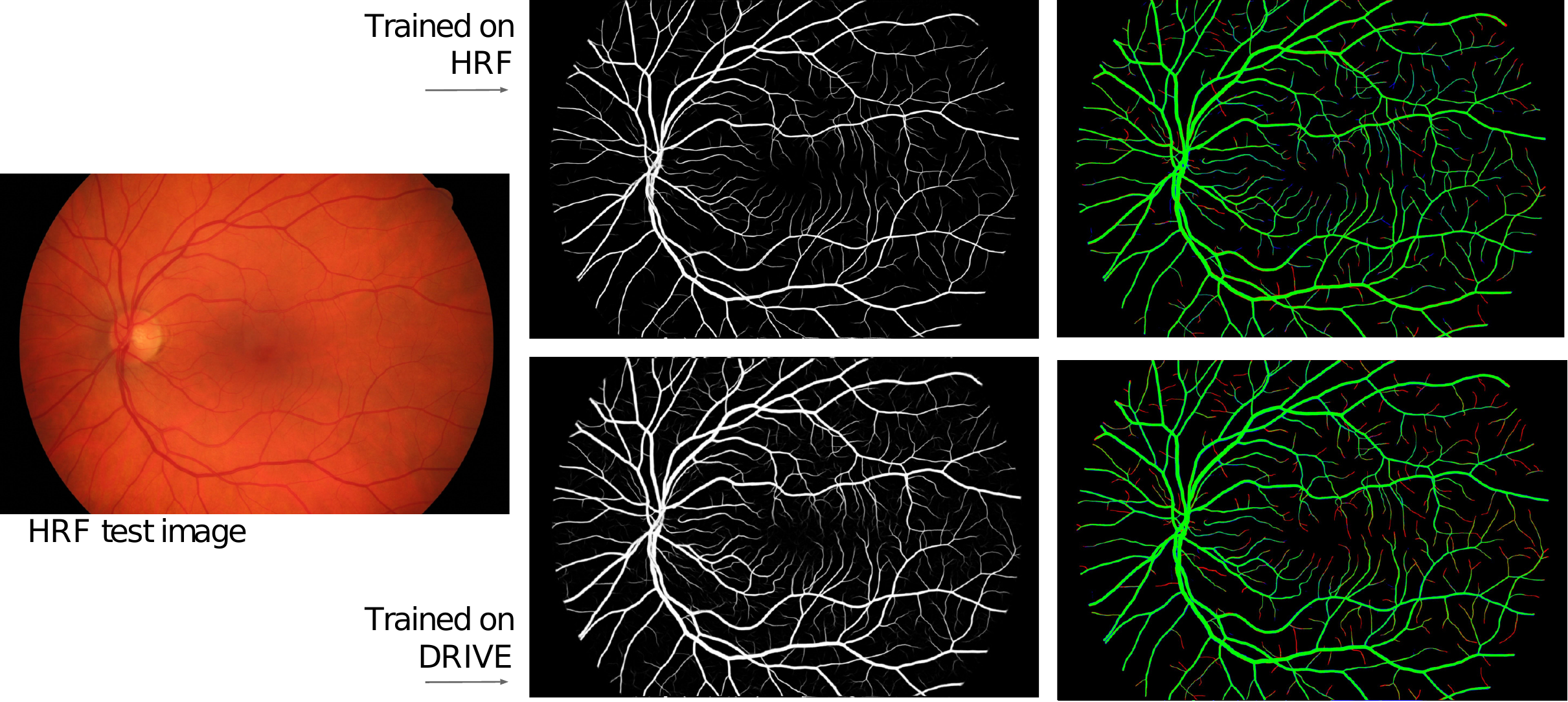}
    \end{center}
    \caption{\small Segmentation result for an HRF test image for a model trained on HRF (upper row) and a model trained on DRIVE (lower row). The second column shows ground truth in white and the third column in green the correctly segmented vessels (zoom in for details).}
    \label{fig:seg}
\end{figure*}

Fig.~\ref{fig:seg} shows example segmentations for an image from the HRF dataset segmented with two models: one trained on the same HRF dataset and another trained on the DRIVE dataset. First, we note that segmentation performance of the two models is comparable (F1: 88.88\% vs 86.98\%). Second, and most importantly, the model trained on DRIVE actually was trained with much lower resolution images, but is still able to generalize to the image material of HRF by making use of our multi-patch, multi-scale framework.

\section{Conclusion}
Our experiments on retinal vessel segmentation have shown that a hybrid framework with large patch sizes extracted at multiple image scales is able to achieve state-of-the-art performance. The combination of these two approaches is able to leverage contextual, large-range image dependencies while at the same time preserving short-range, high-resolution image information. In addition, the framework can better generalize across different datasets as it avoids overfitting on a specific training set's visual appearance.

The overall network structure consists of a simple encoder-decoder structure with ResNet blocks, which provide highly efficient performance at reduced parameter count compared to a U-Net architecture, for example. Our experiments have demonstrated that it possible to reduce the number of parameters even further at minimal loss of segmentation accuracy. It is possible to provide good performance with only 0.8M parameters, which makes our hybrid approach suitable not only for high-precision segmentation, but also for a wide variety of computation-time-sensitive applications. Future developments will concentrate on further improving segmentation accuracy and optimizations to the network architecture.

%
%
%
%

\bibliographystyle{abbrv}
\bibliography{miccai2020}

\begin{thebibliography}{10}

\bibitem{Abramoff2010}
M.~D. Abr{\`{a}}moff, M.~K. Garvin, and M.~Sonka.
\newblock {Retinal Imaging and Image Analysis}.
\newblock {\em IEEE Rev Biomed Eng}, pages 169--208, 2010.

\bibitem{Alom2019}
M.~Z. Alom, C.~Yakopcic, M.~Hasan, T.~M. Taha, and V.~K. Asari.
\newblock {Recurrent residual U-Net for medical image segmentation}.
\newblock {\em Journal of Medical Imaging}, 6(01):1, 2019.

\bibitem{Cai2019}
X.~Cai and Y.~F. Pu.
\newblock {Flattenet: A simple and versatile framework for dense pixelwise
  prediction}.
\newblock {\em IEEE Access}, 7:179985--179996, 2019.

\bibitem{Cheung2011}
C.~Y.-L. Cheung, Y.~Zheng, W.~Hsu, M.~L. Lee, Q.~P. Lau, P.~Mitchell, J.~J.
  Wang, R.~Klein, and T.~Y. Wong.
\newblock {Retinal vascular tortuosity, blood pressure, and cardiovascular risk
  factors.}
\newblock {\em Ophthalmology}, 118(5):812--818, may 2011.

\bibitem{Chollet2017}
F.~Chollet.
\newblock {Xception: Deep learning with depthwise separable convolutions}.
\newblock {\em Proceedings - 30th IEEE Conference on Computer Vision and
  Pattern Recognition, CVPR 2017}, 2017-January:1800--1807, 2017.

\bibitem{Fraz2012}
M.~M. Fraz, P.~Remagnino, A.~Hoppe, B.~Uyyanonvara, A.~R. Rudnicka, C.~G. Owen,
  and S.~A. Barman.
\newblock {An ensemble classification-based approach applied to retinal blood
  vessel segmentation}.
\newblock {\em IEEE Transactions on Biomedical Engineering}, 59(9):2538--2548,
  2012.

\bibitem{Fu2016}
H.~Fu, Y.~Xu, and J.~Liu.
\newblock {DeepVessel: Retinal Vessel Segmentation via Deep Learning and
  Conditional Random Field Huazhu}.
\newblock (December 2017), 2016.

\bibitem{Gharabaghi2013}
S.~Gharabaghi, S.~Daneshvar, and M.~H. Sedaaghi.
\newblock {Retinal image registration using geometrical features}.
\newblock {\em Journal of Digital Imaging}, 26(2):248--258, 2013.

\bibitem{He2015}
K.~He, X.~Zhang, S.~Ren, and J.~Sun.
\newblock {Deep Residual Learning for Image Recognition}.
\newblock {\em Arxiv.Org}, 7(3):171--180, 2015.

\bibitem{Hervella2018}
{\'{A}}.~S. Hervella, J.~Rouco, J.~Novo, and M.~Ortega.
\newblock {Multimodal registration of retinal images using domain-specific
  landmarks and vessel enhancement}.
\newblock {\em Procedia Computer Science}, 126:97--104, 2018.

\bibitem{Howard2017}
A.~G. Howard, M.~Zhu, B.~Chen, D.~Kalenichenko, W.~Wang, T.~Weyand,
  M.~Andreetto, and H.~Adam.
\newblock {MobileNets: Efficient Convolutional Neural Networks for Mobile
  Vision Applications}.
\newblock 2017.

\bibitem{Hubbard1999}
L.~D. Hubbard, R.~J. Brothers, W.~N. King, L.~X. Clegg, R.~Klein, L.~S. Cooper,
  A.~R. Sharrett, M.~D. Davis, and J.~Cai.
\newblock {Methods for evaluation of retinal microvascular abnormalities
  associated with hypertension/sclerosis in the Atherosclerosis Risk in
  Communities Study.}
\newblock {\em Ophthalmology}, 106(12):2269--2280, dec 1999.

\bibitem{Jin2019}
Q.~Jin, Z.~Meng, T.~D. Pham, Q.~Chen, L.~Wei, and R.~Su.
\newblock {DUNet: A deformable network for retinal vessel segmentation}.
\newblock {\em Knowledge-Based Systems}, 178:149--162, 2019.

\bibitem{Kingma2014}
D.~P. Kingma and J.~Ba.
\newblock {Adam: A Method for Stochastic Optimization}.
\newblock pages 1--15, 2014.

\bibitem{Kohler2013}
T.~K{\"{o}}hler, A.~Budai, M.~F. Kraus, J.~Odstr{\v{c}}ilik, G.~Michelson, and
  J.~Hornegger.
\newblock {Automatic no-reference quality assessment for retinal fundus images
  using vessel segmentation}.
\newblock {\em Proceedings - IEEE Symposium on Computer-Based Medical Systems},
  pages 95--100, 2013.

\bibitem{Laibacher2018}
T.~Laibacher, T.~Weyde, and S.~Jalali.
\newblock {M2U-Net: Effective and Efficient Retinal Vessel Segmentation for
  Resource-Constrained Environments}.
\newblock 2018.

\bibitem{Li2016}
Q.~Li, B.~Feng, L.~Xie, P.~Liang, H.~Zhang, and T.~Wang.
\newblock {A cross-modality learning approach for vessel segmentation in
  retinal images}.
\newblock {\em IEEE Transactions on Medical Imaging}, 35(1):109--118, 2016.

\bibitem{Liskowski2016}
P.~Liskowski and K.~Krawiec.
\newblock {Segmenting Retinal Blood Vessels with Deep Neural Networks}.
\newblock {\em IEEE Transactions on Medical Imaging}, 35(11):2369--2380, 2016.

\bibitem{Maninis2016}
K.-k. Maninis, J.~Pont-tuset, P.~Arbel, and L.~V. Gool.
\newblock {Deep Retinal Image Understanding}.
\newblock 2016.

\bibitem{Marin2011}
D.~Mar{\'{i}}n, A.~Aquino, M.~E. Geg{\'{u}}ndez-Arias, and J.~M. Bravo.
\newblock {A new supervised method for blood vessel segmentation in retinal
  images by using gray-level and moment invariants-based features}.
\newblock {\em IEEE Transactions on Medical Imaging}, 30(1):146--158, 2011.

\bibitem{Marino2006}
C.~Mari{\~{n}}o, M.~G. Penedo, M.~Penas, M.~J. Carreira, and F.~Gonzalez.
\newblock {Personal authentication using digital retinal images}.
\newblock {\em Pattern Analysis and Applications}, 9(1):21--33, 2006.

\bibitem{Oliveira2018}
A.~Oliveira, S.~Pereira, and C.~A. Silva.
\newblock {Retinal vessel segmentation based on Fully Convolutional Neural
  Networks}.
\newblock {\em Expert Systems with Applications}, 112:229--242, 2018.

\bibitem{Orlando2017}
J.~I. Orlando, E.~Prokofyeva, and M.~B. Blaschko.
\newblock {A Discriminatively Trained Fully Connected Conditional Random Field
  Model for Blood Vessel Segmentation in Fundus Images}.
\newblock {\em IEEE Transactions on Biomedical Engineering}, 64(1):16--27,
  2017.

\bibitem{Otsu1979}
N.~Otsu.
\newblock {THRESHOLD SELECTION METHOD FROM GRAY-LEVEL HISTOGRAMS.}
\newblock {\em IEEE Trans Syst Man Cybern}, SMC-9(1):62--66, 1979.

\bibitem{Ricci2007}
E.~Ricci and R.~Perfetti.
\newblock {Retinal blood vessel segmentation using line operators and support
  vector classification}.
\newblock {\em IEEE Transactions on Medical Imaging}, 26(10):1357--1365, 2007.

\bibitem{Ronneberger2015}
O.~Ronneberger, P.~Fischer, and T.~Brox.
\newblock {U-Net: Convolutional Networks for Biomedical Image Segmentation}.
\newblock pages 1--8, 2015.

\bibitem{Sandler2018}
M.~Sandler, A.~Howard, M.~Zhu, A.~Zhmoginov, and L.~C. Chen.
\newblock {MobileNetV2: Inverted Residuals and Linear Bottlenecks}.
\newblock {\em Proceedings of the IEEE Computer Society Conference on Computer
  Vision and Pattern Recognition}, pages 4510--4520, 2018.

\bibitem{Shi2016}
W.~Shi, J.~Caballero, F.~Huszar, J.~Totz, A.~P. Aitken, R.~Bishop, D.~Rueckert,
  and Z.~Wang.
\newblock {Real-Time Single Image and Video Super-Resolution Using an Efficient
  Sub-Pixel Convolutional Neural Network}.
\newblock {\em Proceedings of the IEEE Computer Society Conference on Computer
  Vision and Pattern Recognition}, 2016-December:1874--1883, 2016.

\bibitem{Soares2006}
J.~V. Soares, J.~J. Leandro, R.~M. Cesar, H.~F. Jelinek, and M.~J. Cree.
\newblock {Retinal vessel segmentation using the 2-D Gabor wavelet and
  supervised classification}.
\newblock {\em IEEE Transactions on Medical Imaging}, 25(9):1214--1222, 2006.

\bibitem{Son2017}
J.~Son, S.~J. Park, and K.-H. Jung.
\newblock {Retinal Vessel Segmentation in Fundoscopic Images with Generative
  Adversarial Networks}.
\newblock 2017.

\bibitem{Staal2004}
J.~Staal, M.~D. Abr{\`{a}}moff, M.~Niemeijer, M.~A. Viergever, and B.~{Van
  Ginneken}.
\newblock {Ridge-based vessel segmentation in color images of the retina}.
\newblock {\em IEEE Transactions on Medical Imaging}, 23(4):501--509, 2004.

\bibitem{Tajbakhsh2020}
N.~Tajbakhsh, B.~Lai, S.~P. Ananth, and X.~Ding.
\newblock {ErrorNet: Learning Error Representations from Limited Data to
  Improve Vascular Segmentation}.
\newblock In {\em Proceedings - International Symposium on Biomedical Imaging},
  volume 2020-April, pages 1364--1368, 2020.

\bibitem{Tian2010}
J.~Tian, K.~Deng, J.~Zheng, X.~Zhang, X.~Dai, and M.~Xu.
\newblock {Retinal fundus image registration via vascular structure graph
  matching}.
\newblock {\em International Journal of Biomedical Imaging}, 2010, 2010.

\bibitem{Wu2019}
Z.~Wu, C.~Shen, and A.~van~den Hengel.
\newblock {Wider or Deeper: Revisiting the ResNet Model for Visual
  Recognition}.
\newblock {\em Pattern Recognition}, 90:119--133, 2019.

\bibitem{Xiao2018}
B.~Xiao, H.~Wu, and Y.~Wei.
\newblock {Simple baselines for human pose estimation and tracking}.
\newblock {\em Lecture Notes in Computer Science (including subseries Lecture
  Notes in Artificial Intelligence and Lecture Notes in Bioinformatics)}, 11210
  LNCS:472--487, 2018.

\bibitem{Yan2018}
Z.~Yan, X.~Yang, and K.~T. Cheng.
\newblock {Joint segment-level and pixel-wise losses for deep learning based
  retinal vessel segmentation}.
\newblock {\em IEEE Transactions on Biomedical Engineering}, 65(9):1912--1923,
  2018.

\bibitem{You2011}
X.~You, Q.~Peng, Y.~Yuan, Y.~M. Cheung, and J.~Lei.
\newblock {Segmentation of retinal blood vessels using the radial projection
  and semi-supervised approach}.
\newblock {\em Pattern Recognition}, 44(10-11):2314--2324, 2011.

\bibitem{Zhuang2018}
J.~Zhuang.
\newblock {LadderNet: Multi-path networks based on U-Net for medical image
  segmentation}.
\newblock pages 2--5, 2018.

\bibitem{Zhuang2019}
J.~Zhuang, Z.~Chen, J.~Zhang, D.~Zhang, and Z.~Cai.
\newblock {Domain adaptation for retinal vessel segmentation using asymmetrical
  maximum classifier discrepancy}.
\newblock {\em ACM International Conference Proceeding Series}, 2019.

\end{thebibliography}

\end{document}